\documentclass[superscriptaddress,preprint,nofootinbib,amssymb,aps]{revtex4}

\usepackage[intlimits]{amsmath}

\usepackage{graphicx}
\usepackage{dcolumn}
\usepackage{bm}
\usepackage[english]{babel}

\usepackage[colorlinks=true,linkcolor=blue, urlcolor= magenta, citecolor = 
magenta]{hyperref}

\begin{document}

\title{Non-local imprints of gravity on quantum theory}

\author{Michael~Maziashvili}
\email{maziashvili@iliauni.edu.ge}
\affiliation{School of Natural Sciences and Medicine, Ilia State University,\\
3/5 Cholokashvili Ave., Tbilisi 0162, Georgia}

\author{Zurab K. Silagadze}
\email{silagadze@inp.nsk.su}
\affiliation{Budker Institute of Nuclear Physics and Novosibirsk State 
University, Novosibirsk 630 090, Russia}


\begin{abstract}
During the last two decades or so much effort has been devoted 
to the discussion of quantum mechanics (QM) that in some way incorporates 
the notion of a minimum length. This upsurge of research has been prompted by 
the modified uncertainty relation brought about in the framework of string 
theory. In general, the implementation of minimum length in QM can be done 
either by modification of position and momentum operators or by restriction 
of their domains. In the former case we have the so called soccer-ball problem 
when the naive classical limit appears to be drastically different from the 
usual one. Starting with the latter possibility, an alternative approach was 
suggested in the form of a band-limited QM. However, applying momentum cutoff 
to the wave-function, one faces the problem of incompatibility with the 
Schr\"{o}dinger equation. One can overcome this problem in a natural fashion by 
appropriately modifying Schr\"{o}dinger equation. But incompatibility takes 
place for boundary conditions as well. Such wave-function cannot have any more 
a finite support in the coordinate space as it simply follows from 
the Paley-Wiener theorem. Treating, for instance, the simplest 
quantum-mechanical problem of a particle in an infinite potential well, one 
can no longer impose box boundary conditions. In such cases, further 
modification of the theory is in order. We propose a non-local modification of 
QM, which has close ties to the band-limited QM, but does not require a hard 
momentum cutoff. In the framework of this model, one can easily work out 
the corrections to various processes and discuss further the semi-classical 
limit of the theory. 

\end{abstract}

\maketitle
\tableofcontents

\section{Introduction}
General relativity and quantum mechanics are two pillars of modern physics.
However, in their foundational concepts, the two theories differ significantly,
and it is not an easy task to find a unifying framework for them. In general
relativity, basic observables are space-time distances between events, the 
events being intersection points of the world-lines of two objects. Therefore, 
it is assumed that such crossing points are accurately localized in space-time.
This sharp localization is not possible in quantum theory and thus quantum 
mechanics denies the observability of basic events of general relativity 
\cite{1}.

This observation by E.P. Wigner shows the subtlety of the coexistence of 
quantum mechanics and general (or even special) theory of relativity, but does 
not exclude such a possibility. After all, relativistic quantum field theory, 
which combines the principles of special relativity and quantum mechanics, is 
one of our best scientific theories. In fact ``the framework of special 
relativity plus quantum mechanics is so rigid that it practically forces 
quantum field theory upon us'' \cite{2}.

In the case of gravity (general relativity) it is believed that the best
candidate for combining it with quantum mechanics is string theory \cite{3,4}.
String theory is a generalization of quantum field theory in which space-time
emerges from something deeper and more fundamental. However, at present, ``we 
still  don't  know  where  all  these  ideas  are coming from -- or heading 
to ... without  anyone  really  understanding what is behind it'' \cite{2}.

Another approach that tries to combine quantum mechanics and
general relativity is loop quantum gravity \cite{4A,4B}. String theory and 
loop quantum gravity are often viewed as competing, mutually exclusive 
theories. However, it may happen that they are simply highlighting different 
aspects of quantum theories of gravity \cite{4B}, and therefore it is 
worthwhile to pursue both of these paths in the enigmatic and mysterious 
land of quantum gravity.

Since a definitive theory of quantum gravity is still unknown, many 
phenomenological models have been proposed that incorporate the minimum 
observable length predicted by both string theory and quantum gravity models. 
Such models include doubly special relativity \cite{4C} and models with the
generalized uncertainty principle (GUP) \cite{4D,4E,4AA,4BB}. 

A priory, one cannot exclude the possibility that the gravitational field will 
remain classical even at the fundamental level, and only the matter fields 
are quantized. In this case, gravity induces a natural nonlinear and non-local
modification of Schr\"{o}dinger equation \cite{5,6}.  The starting point is
to replace the classical energy-momentum tensor in Einstein's equations by 
the expectation value of the corresponding quantum energy-momentum tensor in 
a given quantum state $\Psi$:
\begin{equation}
R_{\mu\nu}+\frac{1}{2}g_{\mu\nu}R=\frac{8\pi G}{c^4}\langle \Psi|{\hat T}_{\mu\nu}
|\Psi\rangle. 
\label{eq1}
\end{equation}
In the non-relativistic Newtonian limit ${\hat T}_{00}=c^2\hat\rho=
mc^2{\hat \psi}^+\hat\psi$ gives the dominant contribution to the 
energy-momentum tensor, and (\ref{eq1}) becomes the Poisson equation for the 
potential $V({\bf r},t)$
\cite{5,6}:
\begin{equation}
\Delta V=4\pi Gm\,\langle{\hat \psi}^+\hat\psi\rangle,
\label{eq2}
\end{equation}
which can be solved using the well-known expression for the corresponding
Green's function. If now this Newtonian potential energy $mV({\bf r},t)$ is 
introduced in the Schr\"{o}dinger equation, we end up by a nonlinear 
integro-differential equation (the Schr\"{o}dinger-Newton equation), which  
for a one free point-like object takes the form \cite{5,6}
\begin{equation}
i\hbar\partial_t\psi({\bf r},t)=-\frac{\hbar^2}{2m}\Delta \psi({\bf r},t)-
\left (Gm^2\int\frac{|\psi({\bf \xi},t)|^2}{|{\bf \xi}-{\bf r}|}\,d^3\xi
\right )\psi({\bf r},t).
\label{eq3}
\end{equation}

Therefore, it is expected that gravity will induce non-local and nonlinear
modification of the Schr\"{o}dinger equation. However, Schr\"{o}dinger-Newton 
equation suffers from superluminal effects \cite{6}, as all nonlinear 
deterministic generalizations of the Schr\"{o}dinger equation  do \cite{6,7}.
It, at the most, can pretend to be a low-energy effective approximation of a
more fundamental theory. Since the exact form of this more fundamental theory 
(quantum gravity) is not known at present, it seems appropriate to explore 
other approaches also to the question of how the Schr\"{o}dinger equation is 
modified in the presence of gravity.

With the introduction of gravity, one way or another, into quantum theory, 
it becomes fairly clear that at distances comparable to the quantum gravity 
length scale, Planck length, $l_P = (\hbar G_N/c^3)^{1/2}\approx 10^{-33}$\,cm, 
one must drop the standard picture of space-time as a continuum endowed with 
a certain intrinsic geometric structure \cite{Garay:1994en}. These ideas have 
been around for a long time now. Already in 1950s, Wheeler observed that the 
scale dependence of the gravitational action implies large fluctuations of the 
metric and even of the topology on Planck length scale 
\cite{Wheeler:1955zz,Wheeler:1957mu,Wheeler:1964qna}. 

One is thus led to a picture of foamy space-time implying that space-time is 
basically flat on large length scales but is highly curved with all possible 
topologies on the Planck length scale. For instance, the foamy space-time can 
be described in terms of a gas consisting of virtual (Planck size) black holes
--- continually appearing and disappearing \cite{Hawking:1979zw}. 

Apart from this approach, the micro-structure of space-time can be modeled in 
a number of ways: one can represent space-time coordinates by the non-commuting
operators \cite{Snyder:1946qz}, or, equally well, one can assume some sort of 
discrete structure from the very outset \cite{Blokhintsev,Bombelli:1987aa}.  
Keeping aside details and proper mathematical structures related to various 
approaches, the effect of quantum fluctuations of the gravitational field (or 
of space-time geometry) is conventionally summarized either as a source of 
absolute minimum uncertainty in length, which implies smoothing out of 
point-like objects, or as a mechanism for providing momentum cutoff of the 
order of $\hbar/l_P$ to regularize the ultraviolet divergences. Of course, this
may not be a hard momentum cutoff but rather certain modifications of 
dispersion relations of particles that may render loop Feynman graphs 
convergent. 

In view of the sampling theorem in information theory 
\cite{brillouin2004science}, which tells one how to digitize an analog signal 
in a precise way, it was noticed in \cite{Kempf:2010rx} that the application 
of hard momentum cutoff to the fields implies their representation on the 
lattice and thus may be considered as one of the simplest ways for introducing 
discrete space. Inspired by this idea, some basic features of QM with hard 
momentum cutoff has been worked out in \cite{Sailer:2013yk,Sailer:2014hea}. So 
far fairly little attention has been paid to this discussion. First we attempt 
further elaboration of physical and mathematical aspects of such theory. As far
as the classical limit is concerned, the result, if the limit is naively 
applied, turns out to be incompatible with reality. 

Next we proceed to propose somewhat similar non-local model, which, however, 
may be adjusted in such a way as to offer a trivial solution of the 
soccer-ball problem. Corrections due to non-locality are of the same order as 
obtained in various minimum-length deformed models of QM and can be worked out 
without much trouble.            

\section{QM in Hilbert space with hard momentum cutoff}
\subsection{Setting up the basic formalism}
We first put the wave function into a momentum cutoff representation
\begin{eqnarray}\label{Abschneiden}
\psi(x) \,=\, \frac{1}{\sqrt{2\pi\hbar}} \int_{-\text{\ss}}^{\text{\ss}} \mathrm{d}p 
\, \chi(p)\mathrm{e}^{ipx/\hbar} ~, 
\end{eqnarray} where it is understood that the scale $\hbar/\text{\ss}$ is 
related to the minimum length. 
In QM the mean square deviation of coordinate, that is the position 
uncertainty, is usually considered 
as a standard measure of the spread of a wave function in position space. 
It can be shown rigorously 
that the position uncertainty of a wave function with momentum cutoff,  
Eq.\eqref{Abschneiden}, is bounded from below by $\hbar/4\text{\ss}$. To see 
it, we first assume (with no loss of generality) that 
$\langle x \rangle =0$. By using the result 
$$\int_{-\text{\ss}}^{\text{\ss}} \mathrm{d}p \, \left|\frac{\mathrm{d}\chi^*(p)}
{\mathrm{d}p}\right|^2 \,=\,\frac{1}{\hbar^2} 
\int_{-\infty}^\infty \mathrm{d}x \,  x^2 |\psi(x)|^2 ~, $$
which readily follows from the Parseval's formula

$$\int_{-\infty}^\infty \mathrm{d}x \, |\psi(x)|^2 \,=\, \int_{-\text{\ss}}^{\text{\ss}} 
\mathrm{d}p \, |\chi(p)|^2 ~, $$ 
and the following inequalities (the latter one is the Schwarz's 
inequality)  
$$ 2\left|\chi(p)\frac{\mathrm{d}\chi^*(p)}{\mathrm{d}p}\right| \,\geq \, 
\chi(p)\frac{\mathrm{d}\chi^*(p)}{\mathrm{d}p} + 
\chi^*(p)\frac{\mathrm{d}\chi(p)}{\mathrm{d}p} = \frac{\mathrm{d}|\chi(p)|^2}
{\mathrm{d}p} ~, $$ $$ 
\left(\int_{-\text{\ss}}^{\text{\ss}} \mathrm{d}p \, \left|\chi(p)\frac{\mathrm{d}
	\chi^*(p)}{\mathrm{d}p}\right|\right)^2 \,\leq\, 
\int_{-\text{\ss}}^{\text{\ss}} \mathrm{d}p \, \left|\chi(p)\right|^2
\int_{-\text{\ss}}^{\text{\ss}} \mathrm{d}p \, 
\left|\frac{\mathrm{d}\chi^*(p)}{\mathrm{d}p}\right|^2 ~, $$ 
one obtains at once \cite{Price}
$$ 1 = 	\int_{-\text{\ss}}^{\text{\ss}} \mathrm{d}p \, |\chi(p)|^2 \,=\, 	
\int_{-\text{\ss}}^{\text{\ss}} \mathrm{d}p \int_{-\text{\ss}^-}^{p}
\mathrm{d}\xi \, \frac{\mathrm{d}|\chi(\xi)|^2}{\mathrm{d}\xi}\leq $$ $$
2\int_{-\text{\ss}}^{\text{\ss}} \mathrm{d}p \int_{-\text{\ss}}^{p}
\mathrm{d}\xi \, \left|\chi(\xi)\frac{\mathrm{d}\chi^*(\xi)}{\mathrm{d}\xi}
\right| \, \leq \, 2\int_{-\text{\ss}}^{\text{\ss}} \mathrm{d}p 
\int_{-\text{\ss}}^{\text{\ss}}\mathrm{d}\xi \, \left|\chi(\xi)
\frac{\mathrm{d}\chi^*(\xi)}{\mathrm{d}\xi}\right|\leq 
$$ $$ 2\int_{-\text{\ss}}^{\text{\ss}} \mathrm{d} p \, 
\left(\int_{-\text{\ss}}^{\text{\ss}} \mathrm{d}\xi \, 
\left|\chi(\xi)\right|^2\int_{-\text{\ss}}^{\text{\ss}} \mathrm{d}\xi \, 
\left|\frac{\mathrm{d}\chi^*(\xi)}{\mathrm{d}\xi}
\right|^2\right)^{1/2} =
\frac{4\text{\ss}}{\hbar} \left(\int_{-\infty}^{\infty} \mathrm{d}x\, x^2 
|\psi(x)|^2 \right)^{1/2}. $$

Let us note that the wave functions of the form \eqref{Abschneiden} obey the 
integral equation    
\begin{eqnarray}\label{integralgleichung} &
\psi(x) = \frac{1}{2\pi\hbar} \int\limits_{-\text{\ss}}^{\text{\ss}} \mathrm{d}p \, 
\mathrm{e}^{ipx/\hbar} \int\limits_{-\infty}^\infty \mathrm{d}y \, 
\mathrm{e}^{-ipy/\hbar}\psi(y) =
\frac{1}{\pi} \int\limits_{-\infty}^\infty \mathrm{d}y \, \psi(y) \,
\frac{\sin\big[\text{\ss} (x-y)/\hbar\big]}{x-y}. &
\end{eqnarray} Thus the Schr\"{o}dinger equation is now supplemented by this 
integral one. But it is plain to note that 
this system of equations does not always admit a solution even in simple cases.
One may easily observe that in general
the initial state given by Eq.\eqref{Abschneiden} will evolve into the 
function which does not admit this sort of representation. 
It suffices to consider an infinitesimal time development           
\begin{eqnarray}\label{zeitentwicklung} &&
\psi(t, x) \,=\, \psi_0(x) -\frac{it}{\hbar} \widehat{H} \psi_0(x) + O
\left(t^2\right) \,=\,  \nonumber \\ &&
\psi_0(x) +\frac{it\hbar}{2m}
\psi_0^{''}(x) -\frac{it}{\hbar} V(x)\psi_0(x)+ O\left(t^2\right)~. 
\end{eqnarray} The problem in Eq.\eqref{zeitentwicklung} arises because of the 
term $V(x)\psi_0(x)$. Even if both $V(x)$ and 
$\psi_0(x)$ were taken to have the same compact support in momentum space, the 
product $V(x)\psi_0(x)$ will not have 
the same support in general. The significance of this fact in the context 
similar to our discussion 
was emphasized in \cite{Elze:2013mpa}.        

Thus, one needs to reformulate the set up in a more consistent way. The 
operator entering the Eq.\eqref{integralgleichung} 
\begin{eqnarray}\label{projektionsoperator}
\widetilde{\psi}(x)  \,=\, \frac{1}{\pi} \int_{-\infty}^\infty \mathrm{d}y \, 
\psi(y) \, \frac{\sin\big[\text{\ss} (x-y)/\hbar\big]}{x-y} ~,
\end{eqnarray} has the property that it projects out of $\psi(x)$ the part 
$\widetilde{\psi}(x)$ - the Fourier transform 
of which coincides with that of $\psi$ in $|p| < \text{\ss}$ and vanishes 
elsewhere. By using the integral representation 
$$\frac{1}{2\pi\hbar} \int_{-\text{\ss}}^{\text{\ss}} \mathrm{d}p \, 
\mathrm{e}^{ip(x - y)/\hbar}   \,=\,   \frac{1}{\pi} \, 
\frac{\sin\big[\text{\ss} (x-y)/\hbar\big]}{x-y} ~, $$ 
it is easy to verify that   
$$\frac{1}{\pi^2} \int_{-\infty}^{\infty}\mathrm{d}y \, \frac{\sin\big[\text{\ss} 
	(x-y)/\hbar\big]}{x-y}  \times   
\frac{\sin\big[\text{\ss} (y-z)/\hbar\big]}{y-z} \,=\,   \frac{1}{\pi} \, 
\frac{\sin\big[\text{\ss} (x-z)/\hbar\big]}{x-z} ~. $$ 
This property is typical for projector operators: 
$\widehat{\Pi}^2 =\widehat{\Pi}$. A self-consistent treatment 
of the problem can be obtained by considering a non-local modification of 
the Schr\"{o}dinger equation 

\begin{eqnarray}\label{erste}
i\hbar \partial_t \psi(t, x) \,=\,  \frac{1}{\pi} \int_{-\infty}^\infty \mathrm{d}y
\,  \frac{\sin\big[\text{\ss} (x-y)/\hbar\big]}{x-y} 
\left\{ -\frac{\hbar^2}{2m}\frac{\mathrm{d}^2}{\mathrm{d}y^2} \,+\, V(y) 
\right\}\psi(t, y)  ~, 
\end{eqnarray} or more minimalistic version
\begin{eqnarray}\label{zweite}
i\hbar \partial_t \psi(t, x) \,=\, -\frac{\hbar^2}{2m}\frac{\mathrm{d}^2
	\psi(t, x)}{\mathrm{d}x^2} \,+\,     \frac{1}{\pi} 
\int_{-\infty}^\infty \mathrm{d}y \,  \frac{\sin\big[\text{\ss} (x-y)/\hbar\big]}
{x-y} \, V(y) \psi(t, y)   ~. 
\end{eqnarray} In itself, the purpose of modification of the kinetic term is 
not clear, as the minimalistic version, Eq.\eqref{zweite}, 
already guaranties that the solution will be of the form \eqref{Abschneiden} 
under assumption that initial state has such a form. 
It is plain to see that if the scalar product is defined in the standard manner 
$$\langle \psi_1 |\psi_2\rangle \,=\, \int_{-\infty}^\infty\mathrm{d}x \, 
\psi_1^*(x)\psi_2(x) ~, $$ 
then in Eq.\eqref{erste} the modified kinetic term is Hermitian 
while the potential one is not. 
The potential term in Eq.\eqref{zweite} has the same problem. Accordingly, the 
potential term should be modified in such a way as to make the Hamiltonian 
Hermitian. The Hermiticity is recovered by the following redefinition of this 
term  \cite{Sailer:2013yk}
\begin{eqnarray}\label{hermitisch1}
\frac{1}{\pi} \int_{-\infty}^\infty \mathrm{d}y \,  \frac{\sin\big[\text{\ss} 
	(x-y)/\hbar\big]}{x-y} \, \frac{V(y)+V(x)}{2} \, \psi(y) ~. 
\end{eqnarray} Equally well one could use the following redefinition          
\begin{eqnarray}\label{hermitisch2}
\frac{1}{\pi} \int_{-\infty}^\infty \mathrm{d}y \,  \frac{\sin\big[\text{\ss} 
	(x-y)/\hbar\big]}{x-y} \, V\left(\frac{x+y}{2}\right)   \psi(y) ~. 
\end{eqnarray} Before discussing some more technical aspects, let us work out 
the classical limits.    

\subsection{Classical limit}
By using the momentum operator, one can write the Eq.\eqref{hermitisch1} as 
$$ \frac{1}{2\pi}\int_{-\infty}^\infty \mathrm{d}\xi \, \frac{\sin(\text{\ss} 
	\xi /\hbar)}{\xi} \Big\{V(x)+V(x-\xi) \Big\} \, \psi(x-\xi)= 
$$ $$ 	\frac{1}{2\pi}\int_{-\infty}^\infty \mathrm{d}\xi \, 
\frac{\sin(\text{\ss} \xi /\hbar)}{\xi} \Big\{V(x)
\mathrm{e}^{-i\widehat{p}\xi /\hbar} +\mathrm{e}^{-i\widehat{p}\xi /\hbar}V(x)\Big\} \, 
\psi(x) ~. $$ 
In the case of Eq.\eqref{hermitisch2} one obtains 
$$\frac{1}{\pi}\int_{-\infty}^\infty \mathrm{d}\xi \, \frac{\sin(\text{\ss} 
	\xi /\hbar)}{\xi} \mathrm{e}^{-i\widehat{p}\xi /2\hbar}
V(x)\mathrm{e}^{-i\widehat{p}\xi /2\hbar} \psi(x) ~. $$ 
In the classical regime, when the momentum and position 
operators commute,  the potentials in both cases get modified as follows 
\begin{eqnarray}
V(x) \,\rightarrow \,  V(x)\theta\big(\text{\ss} - |p|\big) ~, 
\end{eqnarray} where $\theta$ is a Heaviside function. Thus, the classical 
motion with $|p| > \text{\ss}$ becomes essentially free. 
Basically the same result is obtained in section \ref{klasika}. If one applies 
the Eq.\eqref{erste} in which either 
Eq.\eqref{hermitisch1} or Eq.\eqref{hermitisch2} is used for the potential 
term, then the classical limit will be represented by a modified Hamiltonian   
\begin{eqnarray}
\mathcal{H} \,\rightarrow \,  \mathcal{H}\theta\big(\text{\ss} - |p|\big) ~.
\end{eqnarray} In both cases the result seems obviously incompatible with 
the reality. This so called "Soccer-ball problem" is common to many 
quantum-gravity inspired modifications of quantum mechanics \cite{Sabina}, 
and, perhaps, indicates non-triviality of considering multi-particle states 
and corresponding macroscopic bodies in such theories \cite{Quesne,Tkachuk}.
Therefore,  perhaps  ``composition law problem" would be more 
appropriate terminology. Relativity brings another twist in this composition 
law problem \cite{7A,7B,7C,7D}, which, in our oppinion,  is currently far from 
a final resolution.

\subsection{Incompatibility with box-boundary conditions: An "infinite" 
	    potential well}
One more technical issue that deserves attention is the potential cutoff. 
To elucidate this point let us consider the problem of infinite potential 
well. The equation determining energy levels can be written as   
$$ \mathcal{E}\psi \,=\, 	-\, \frac{\hbar^2}{2m}\frac{\mathrm{d}^2\psi(x)}
{\mathrm{d}x^2} \,+\, \frac{1}{2\pi}\int_{-\infty}^\infty\mathrm{d}\xi \, 
\frac{\sin(2\text{\ss}(x-\xi)/\hbar)}{x \,-\, \xi} \, V(\xi)\psi(2\xi - x)  
~, $$  
where we have used Eq.\eqref{hermitisch2} and instead of 
variable $y$ introduced a new variable $\xi$: 
$y = 2\xi -x$. One can consider this equation for a finite well
\begin{eqnarray}
V(x) \,=\, \begin{cases}
0       & \text{for } ~ -l < x <l \\
V_0 & \text{for } ~ |x| \geq l\\
\end{cases}
\end{eqnarray} and then let $V_0 \to \infty$. Hence, one obtains 
\begin{eqnarray}\label{eigenwertproblem}
-\, \frac{\hbar^2}{2m}\frac{\mathrm{d}^2\psi(x)}{\mathrm{d}x^2} \,+\,  
\frac{V_0}{2\pi}\int_{-\infty}^{-l}\mathrm{d}\xi \, 
\frac{\sin(2\text{\ss}(x-\xi)/\hbar)}{x \,-\, \xi} \, \psi(2\xi - x)     
\nonumber  \\ \,+\,  \frac{V_0}{2\pi}\int_{l}^{\infty}\mathrm{d}\xi \, 
\frac{\sin(2\text{\ss}(x-\xi)/\hbar)}{x \,-\, \xi} \, \psi(2\xi - x) \,=\,  
\mathcal{E}\psi ~. ~~~~ 
\end{eqnarray} Letting $V_0 \to \infty$, one has to impose that 
the wave-function vanishes outside the well - in order to avoid infinite 
terms. That implies the equation   
$$ -\, \frac{\hbar^2}{2m}\frac{\mathrm{d}^2\psi(x)}{\mathrm{d}x^2} \,=\,  
\mathcal{E}\psi ~, ~\psi = 0 ~\text{for}~ |x|\geq l  ~, $$  
which can be solved in a standard manner. But, of course, 
in this case the solution will not have the form 
\eqref{Abschneiden} since the momentum cutoff prevents the function from 
having a finite support in the position space. 
In view of the Paley-Wiener theorem, such functions are analytic in the 
entire complex plane. As a result, it follows that 
the non-trivial function of the form \eqref{Abschneiden} can not vanish on 
any interval of the $x$-axis \cite{Price,Slepian,Landau}. 
Because of this, one can not impose zero boundary conditions outside the well 
when dealing with an infinite well problem. 
Then it is clear that the limit $V_0\to\infty$ cannot be taken in 
Eq.\eqref{eigenwertproblem} and, therefore, one has to set 
the value of $V_0$ somehow. The only reasonable way for doing this seems to 
be the use of scale $\text{\ss}$. That is, the potential of an infinite well 
could be modified as   
\begin{eqnarray}\label{potentialschacht}
V(x) \,=\, \begin{cases}
0       & \text{for } ~ -l < x <l  ~, \\ 
V_0=\text{\ss}^2/2m & \text{for } ~ |x| \geq l  ~.
\end{cases} 
\end{eqnarray}   

Let us note in passing that the corrections to the low-lying energy levels can 
be found by exploiting the standard perturbation theory, see section 
\ref{shestsorebebi}. On the other hand, for energy levels which are high enough 
(close to $\text{\ss}^2/2m$) one may neglect the second derivative in 
Eq.\eqref{eigenwertproblem}. Under this assumption, one arrives at the 
equation     
$$ \frac{V_0}{2\pi}\left [\int_{-\infty}^{-l}\mathrm{d}\xi \, 
\frac{\sin(2\text{\ss}(x-\xi)/\hbar)}{x \,-\, \xi} \, \psi(2\xi - x)  +
\right . $$ $$ \left .  
\int_{l}^{\infty}\mathrm{d}\xi \, \frac{\sin(2\text{\ss}(x-\xi)/
	\hbar)}{x \,-\, \xi} \, \psi(2\xi - x)\right ] =
\mathcal{E}\psi, $$  
which looks somewhat like the Eigenwertproblem addressed in 
\cite{Slepian}. One could try to use the solution 
from the cited paper and look for the approximate solutions of 
Eq.\eqref{eigenwertproblem}. Certainly, in the end one has 
to check the validity of the applied approximation.    

Guided by the example of infinite well, one may loosely argue that the cutoff 
on the potential should be applied as a general rule. This way, the harmonic 
oscillator gets modified as 
\begin{eqnarray}\label{oszillator}
V(x) \,=\, \begin{cases}
m\omega^2x^2/2      & \text{for } ~ -\text{\ss}/m\omega < x <\text{\ss} 
/m\omega ~, \\
\text{\ss}^2/2m & \text{for } ~ |x| \geq \text{\ss}/m\omega ~.
\end{cases}
\end{eqnarray} 

While in elementary particle physics one may consider various 
thought experiments for arguing that the laws of physics 
forbid us from reaching Planck energy scale \cite{Casher:1995qc,Casher:1997rr},
from the point of view of classical physics this sort of modification of the 
potential is hard to understand. 

In itself, the cutoff on the potential has the following useful role in 
accordance with the  the context of our discussion. 
The standard Schr\"{o}dinger equation with the cutoff potential implies that 
the spread of bound states cannot be made smaller
than $\hbar / \text{\ss}$. One can easily verify this statement by using 
simple examples of "infinite" well and harmonic oscillator. 

\subsection{Relation to the deformed Weyl-Heisenberg algebra}
Instead of $p$ one can introduce a new variable $\mathcal{P}$, which covers 
the whole axis  
\begin{eqnarray*}
	\mathcal{P} \,=\, \frac{2 \text{\ss}}{\pi} \tan\left(\frac{\pi p}
	{2\text{\ss}}\right) ~.  
\end{eqnarray*} 
The Hamiltonian takes the form 
\begin{eqnarray*}
	\widehat{H} \,=\,  \frac{2\text{\ss}^2 \arctan^2\big(\pi 
		\widehat{\mathcal{P}}/ 2\text{\ss}\big)}{\pi^2m} \,+\, 
	V\big(\widehat{\mathcal{X}} \,\big) ~,  
\end{eqnarray*} 
where $\widehat{\mathcal{X}}$  is the position operator with the commutation 
relation
\begin{eqnarray*}
	&&\left[\widehat{\mathcal{X}}, \widehat{\mathcal{P}}\right] \,=\, 
i\hbar \, 
	\left(1 \,+\, \frac{\pi^2 
		\widehat{\mathcal{P}}^2 }{4\text{\ss}^2}\right) ~.  
\end{eqnarray*}

Certainly the above change of variables is not the only admissible one. 
Equally well one could consider 
$$p \,=\,  \text{\ss}  \tanh \big(\mathcal{P}/\text{\ss}\big) ~, $$ 
which leads to
$$ \widehat{H} \,=\,  \frac{\text{\ss}^2 \tanh^2\big( \widehat{\mathcal{P}}/ 
	\text{\ss}\big)}{2m} \,+\, V\big(\widehat{\mathcal{X}} \,\big) ~, $$
with the commutation relations  
$$ \left[\widehat{\mathcal{X}}, \widehat{\mathcal{P}}\right] \,=\, 
\frac{ i\hbar}{1 \,-\, \tanh^2 
	\big(\widehat{\mathcal{P}}/\text{\ss}\big)}  ~. $$ 
One can construct many other examples as well. 

In passing, let us mention that it is straightforward to construct Hilbert 
space representation of the deformed 
Weyl-Heisenberg algebra by using the mapping: $\mathcal{P}=f(p)$. Namely, 
in momentum representation the integration measure is altered as
$$ \mathrm{d}p \rightarrow \mathrm{d}\mathcal{P} \, \frac{\mathrm{d}f^{-1}
	(\mathcal{P})}{\mathrm{d}\mathcal{P}} ~, $$ 
and the scalar product takes the form 
$$ \langle \psi_1(\mathcal{P})|\psi_2(\mathcal{P})\rangle \,=\, 
\int_{-\infty}^\infty \mathrm{d}\mathcal{P} \, 
\frac{\mathrm{d}f^{-1}(\mathcal{P})}{\mathrm{d}\mathcal{P}} \, 
\psi_1^*(\mathcal{P})\psi_2(\mathcal{P}) ~ . $$
The momentum operator multiplies a state by $\mathcal{P}$, 
while the position operator is defined by the replacement 
$$\widehat{x} \,=\, i\hbar\,\frac{\mathrm{d}}{\mathrm{d}p} ~\rightarrow~ 
i\hbar\,f'\Big(f^{-1}(\mathcal{P})\Big)
\frac{\mathrm{d}}{\mathrm{d}\mathcal{P}} ~. $$

Let us note that in the above examples one could leave the Hamiltonian 
unaltered. Then it would mean in $\widehat{x}, 
\widehat{p}$ variables that not only Hilbert space is restricted but 
Hamiltonian also is modified. More precisely, in the 
first example it amounts to the deformation of momentum operator 
$$ \widehat{p} ~\rightarrow ~ \frac{2\text{\ss}}{\pi}\tan\left(\frac{\pi 
	\widehat{p}}{2\text{\ss}}\right) ~, $$
and to the deformation 
$$ \widehat{p} ~\rightarrow ~ \text{\ss} \tanh^{-1}\left(\frac{ \widehat{p}}
{\text{\ss}}\right) ~, $$
in the second example. In both cases  the deformed momentum 
(momentum at the high energy) is not bounded. This is inconsistent with double 
special relativity, in which momentum has an invariant maximum called Planck's 
momentum. The construction of a generalized uncertainty principle with both 
minimum length and maximum momentum is considered, for example, in 
\cite{8A}, while in \cite{8B} a generalization of the uncertainty principle  
is presented that introduces the existence of a maximal observable momentum, 
but does not entail the minimum length.

We note in passing that one could introduce somewhat different deformed 
momentum-operator on the basis of sampling theorem. Namely, the cutoff 
representation of the wave-function \eqref{Abschneiden} implies that 
\cite{brillouin2004science} 
$$\psi(x) \,=\, \sum_{j=-\infty}^\infty\psi\left(\frac{\hbar \pi j}
{\text{\ss}}\right) \, \frac{\sin(\text{\ss} x/\hbar - \pi j)}
{\text{\ss} x/\hbar - \pi j} ~. $$
In other words, such wave-function is determined by knowing its 
values at the points $x_j = \hbar \pi j/\text{\ss}$. 
It "suggests" to replace the derivative with some approximate expression. For 
instance, by using the translation operator 
$\widehat{U}(\delta x)\psi(x) = \psi(x+ \delta x)$, one could introduce the 
deformed momentum as  
$$ \widehat{\mathcal{P}} \,=\, \frac{\text{\ss}}{i\pi} \left(\widehat{U}
(\hbar\pi/2\text{\ss}) \,-\, \widehat{U}(-\hbar\pi/2\text{\ss})  \right) \,=\, 
\frac{2\text{\ss}}{\pi}\sin\left(\frac{\pi \widehat{p}}{2\text{\ss}}\right) ~. 
$$
Somewhat similar discussion can be found in \cite{Elze:2016gft,Elze:2017evx}.

Correspondingly, the deformed Weyl-Heisenberg algebra will take the form 
\begin{equation}
\left[\widehat{\mathcal{X}}, \widehat{\mathcal{P}}\right] \,=\, 
i\hbar\sqrt{1 \,-\, \frac{\pi^2
		\widehat{\mathcal{P}}^2}{\text{\ss}^2}} \,=\,  i\hbar 
\left( 1 \,-\, \frac{\pi^2
	\widehat{\mathcal{P}}^2}{2\text{\ss}^2} \,-\, 
\frac{\pi^4\widehat{\mathcal{P}}^4}{8\text{\ss}^4}  \,-\, \cdots  \right) ~,
\end{equation} 
where $\widehat{\mathcal{X}} =\widehat{x}$. In momentum representation one 
obtains        
$$\widehat{\mathcal{X}} \,=\, i\hbar \sqrt{1 \,-\, \frac{\pi^2\mathcal{P}^2}
	{\text{\ss}^2}} \, \frac{\mathrm{d}}{\mathrm{d} 
	\mathcal{P}} ~, ~~ \widehat{\mathcal{P}} \,=\, \mathcal{P} ~, $$ 
where it is understood that $\mathcal{P} ^2 < 
\text{\ss}^2/\pi^2$. 

\subsection{Brief summary}
Let us briefly summarize the key points concerning band-limited QM. As it was 
suggested in \cite{Sailer:2013yk,Sailer:2014hea}, 
by using the projection operator \eqref{projektionsoperator}, one can easily 
find the modified Schr\"{o}dinger equation compatible 
with the momentum cutoff of the wave function. But the problem of 
compatibility still persists as far as we are concerned with
the boundary conditions. An important point here is that if momentum cutoff 
is imposed on $\psi(x)$. then it cannot be supported 
on a finite interval in the coordinate space. Thus, if we make the potential 
walls impenetrable, then the Schr\"{o}dinger equation will 
not have solution of the form \eqref{Abschneiden}. A characteristic difficulty 
of minimum-length deformed QM is that in 
the semi-classical limit one is not lead to well defined classical picture, 
without a proper treatment of multi-particle macroscopic 
objects. The band limited QM also does not automatically go over to the well 
defined classical theory.       

\section{Non-local QM without hard momentum cutoff}

\subsection{Some introductory remarks}
\label{shesavali}

In the sequel we shall outline a possible non-local generalization of quantum 
mechanics, which has simple logical connections 
to the band-limited QM and to the micro structure of the background space. 
First we observe that the UV cutoff of 
the wave-function can be understood as a spatial averaging 
\begin{eqnarray}\label{Verschmierte Wellenfunktion}  
\widetilde{\Psi}(\mathbf{r}) \,=\, \int\mathrm{d}^3\xi \, f(\xi)\Psi(\mathbf{r}
- \boldsymbol{\xi}) ~,  
\end{eqnarray} where the characteristic size of the test function $f(\xi)$ is 
assumed to be of the order of $l_P$. Namely, using for the sake of simplicity 
a Gaussian test function  

\begin{eqnarray}\label{Gauss-Verteilung}
f(\xi) \,=\, \left(\pi l_P^2\right)^{-3/2}\mathrm{e}^{-\xi^2/l_P^2} ~, 
\end{eqnarray} the Fourier transform of \eqref{Verschmierte Wellenfunktion} 
will take the form 
$$ \widetilde{\chi}(\mathbf{k}) \,\propto\, \mathrm{e}^{-k^2l_P^2/4}
\chi(\mathbf{k}) ~, $$
clearly indicating the (exponential) suppression of the 
Fourier modes: $k^2l_P^2 \gg 1$. Next we observe 
that the averaging of the wave-function can naturally be understood as 
a coarse graining due to grainy structure of the space 
or as a result of background space fluctuations. For instance, one may bear 
the following simple picture in mind. Various 
thought experiments for measuring a background space show that its resolution 
is limited by the Planck length: 
$l_P = (\hbar G_N/c^3)^{1/2}\approx 10^{-33}$\,cm \cite{Garay:1994en}. This fact 
might be taken to suggest that background space undergoes fluctuations in the 
sense that a position of point can not be known precisely but rather 
with some probability. This feature of the background space can be described 
effectively by specifying a distribution 
function $f(\xi)$, so that the integral $\int f(\xi)\mathrm{d}^3\xi$ over some 
region, $l^3$, in the vicinity of any point, can 
be interpreted as the probability that a position of this point is known with 
the precision $l^3$. In other words, that is the 
probability that a given point lies (in the operational sense) within this 
volume. For we are dealing with isotropic and 
homogeneous background space, it is naturally assumed that $f$ depends just 
on $\xi$ and does not depend on $\mathbf{r}$.          

Physically, an introduction of the above distribution function implies that 
one can measure only averaged quantities over a space region. But the 
averaging must be done with same care. In particular, as the Schr\"{o}dinger 
equation involves the product of a potential and a wave function at the same 
point, some care is needed to define the average value of this product 
"properly" in order to ensure the Hermiticity of Hamiltonian. We shall 
discuss these questions in what follows. 

\subsection{Modified Schr\"{o}dinger equation}

Let us start by considering an averaged wave-function 
\eqref{Verschmierte Wellenfunktion}. If $\Psi(\mathbf{r})$ 
satisfies the Schr\"{o}dinger equation, then the equation for 
$\widetilde{\Psi}$ takes the form 

\begin{eqnarray}\label{Nicht-Hermitesche}
i\hbar \partial_t \widetilde{\Psi}(\mathbf{r}) \,=\, -\frac{\hbar^2}{2m} 
\Delta \widetilde{\Psi}(\mathbf{r}) \,+\,  
\int\mathrm{d}^3\xi \, f(\xi)V(\mathbf{r} - \boldsymbol{\xi})\Psi(\mathbf{r} 
- \boldsymbol{\xi}) ~. 
\end{eqnarray} Replacing in the integrand $\Psi$ with $\widetilde{\Psi}$, one 
may naturally interpret this integral as an average value of 
$V\widetilde{\Psi}$. If we define the scalar product in a standard way  
$$ \langle \widetilde{\Psi}_1 | \widetilde{\Psi}_2 \rangle \,=\, \int 
\mathrm{d}^3 \, \widetilde{\Psi}^*_1(\mathbf{r}) 
\widetilde{\Psi}_2(\mathbf{r}) ~, $$
then the Hamiltonian in Eq.\eqref{Nicht-Hermitesche} 
(in which $\Psi$ is replaced by $\widetilde{\Psi}$) 
is clearly non-Hermitian. One can, however, easily modify the 
Eq.\eqref{Nicht-Hermitesche} in such a way as to render 
the Hamilton operator Hermitian. For instance, one can put the modified 
equation in the form (from now on we omit the tilde) 
\begin{eqnarray}\label{Grundgleichung}
i\hbar \partial_t \Psi(\mathbf{r}) \,=\, -\frac{\hbar^2}{2m} \Delta 
\Psi(\mathbf{r}) \,+\,  \int\mathrm{d}^3x' \, f(|\mathbf{r} - 
\mathbf{r}'|)V\left(\frac{\mathbf{r} +\mathbf{r}'}{2}\right)
\Psi(\mathbf{r}') ~. 
\end{eqnarray} In the limit $l_P \to 0$, $f(|\mathbf{r} - \mathbf{r}'|)$ 
tends to $\delta(\mathbf{r} - \mathbf{r}')$ and one 
arrives at the standard Schr\"{o}dinger equation. The last term in 
Eq.\eqref{Grundgleichung} is just a smeared out version 
of the product $V(\mathbf{r})\Psi(\mathbf{r})$. Let us note that this sort 
of equations have been discussed extensively in the context of 
nuclear physics \cite{1962NucPh..32..353P}.     

One more relatively simple modification of the Schr\"{o}dinger equation that 
follows from the above discussion might be 
\begin{eqnarray}\label{einfache Version}
i\hbar \partial_t \Psi(\mathbf{r}) \,=\, -\frac{\hbar^2}{2m} \Delta 
\Psi(\mathbf{r}) \,+\,  \Psi(\mathbf{r})\int\mathrm{d}^3\xi \, 
f(\xi)V(\mathbf{r} - \boldsymbol{\xi}) ~. 
\end{eqnarray} In fact, one could use the Eq.\eqref{einfache Version} for 
estimating gravitational corrections to the quantum 
mechanics, but as it is almost trivial generalization, we will mainly focus 
on Eq.\eqref{Grundgleichung}.  

\subsection{Digression on the averaging as a similarity transformation}

This may be of some conceptual interest to note that the averaging given by 
Eqs.(\ref{Verschmierte Wellenfunktion}, 
\ref{Gauss-Verteilung}) can be viewed
as the similarity transformation \cite{Hohlfeld:1993} 
\begin{equation} \widetilde{\Psi}\,=\, \hat B \Psi ~, ~~
\hat{\widetilde{H}}=\hat B\,\hat H \hat B^{-1} ~, ~~ \text{where}~~\hat B=
\mathrm{e}^{\frac{l_P^2 \boldsymbol{\nabla}^2}{4}} ~ ~\text{and} ~~ f(\mathbf{r})\,=
\mathrm{e}^{\frac{l_P^2}{4}\boldsymbol{\nabla}^2}\delta (\mathbf{r}) ~.
\label{eq4}
\end{equation} This transformation, which can be viewed as a formal analog of 
the Kadanoff-Wilson blocking procedure 
in renormalization theory  \cite{ElHattab:1997gj,Liao:1992fm}, is obviously 
non-unitary and therefore the new Hamiltonian is non-Hermitian. The lack of 
Hermiticity can be interpreted in physical terms as a result of high-frequency 
modes cutoff. The modified Schr\"{o}dinger equation obtained by the 
transformation \eqref{eq4} is non-local 
\begin{eqnarray}
i\hbar \partial_t \widetilde{\Psi}(\mathbf{r}) \,=\, -\frac{\hbar^2}{2m} 
\Delta \widetilde{\Psi}(\mathbf{r}) \,+\,\mathrm{e}^{\frac{l_P^2 
		\boldsymbol{\nabla}^2}{4}}
V(\mathbf{r})\mathrm{e}^{-\frac{l_P^2 \boldsymbol{\nabla}^2}{4}}
\widetilde{\Psi}(\mathbf{r})=  -\frac{\hbar^2}{2m} \Delta \widetilde{\Psi}
(\mathbf{r})\,+\, \nonumber \\
\int  \mathrm{d}^3\xi \, f(\mathbf{r}-\boldsymbol{\xi})V(\boldsymbol{\xi})
\sum\limits_{n=0}^\infty\frac{(-l_P^2/4)^n}{n!}\Delta^n\widetilde{\Psi}
(\boldsymbol{\xi}) ~. ~~~~~~
\label{eq5}
\end{eqnarray} Retaining the scalar product in its standard form  
$$ \langle \widetilde{\Psi}_1 | \widetilde{\Psi}_2 \rangle \,=\, \int 
\mathrm{d}^3 \, \widetilde{\Psi}^*_1(\mathbf{r}) \widetilde{\Psi}_2
(\mathbf{r}) ~, $$
if we want to regain a well defined quantum-mechanical picture, 
we have to modify the Hamiltonian \eqref{eq4} in such a way
that its Hermiticity is restored. We shall not pursue the general 
consideration further, but instead restrict ourselves to the limiting case 
when average momentum is much smaller than $\hbar/l_P$. Under this assumption, 
in \eqref{eq5} one can discard first and higher order terms in 
$\mathbf{p}^2l_P^2$  retaining only zeroth order term. This way one arrives at 
a non-Hermitian Schr\"{o}dinger equation, which is a predecessor of 
Eq.\eqref{Grundgleichung}.

One could provide some further technical details concerning the blocking 
transformation (\ref{Verschmierte Wellenfunktion}, \ref{Gauss-Verteilung}, 
\ref{eq4}). For inverting this transformation, one usually uses 
the solution of the Fredholm-type integral equation 
\eqref{Verschmierte Wellenfunktion} by a Fourier transform method. 
For the Gaussian kernels this may imply ill-posed 
problems due to the presence of a fast growing Gaussian function in the 
deconvolution integral \cite{Ulmer:2003}. However, there exist alternative 
methods of deconvolution of Gaussian kernels, avoiding ill-posed problems 
\cite{Ulmer:2003,Ulmer:2010}. One of such methods is what follows. Let us 
first note that
$$\Psi(\mathbf{r})=e^{-\frac{l_P^2\boldsymbol{\nabla}^2}{4}}\widetilde{\Psi}
(\mathbf{r})=\int \mathrm{d}^3\xi\,\delta(\mathbf{r}-\boldsymbol{\xi})\,
e^{-\frac{l_P^2\boldsymbol{\nabla}^2}{4}}\widetilde{\Psi}(\boldsymbol{\xi})= $$ $$ 
\int \mathrm{d}^3\xi\,\widetilde{\Psi}(\boldsymbol{\xi})\,
e^{-\frac{l_P^2\boldsymbol{\nabla}^2}{4}}\delta(\mathbf{r}-\boldsymbol{\xi}) ~. $$  
On the other hand,
$$ e^{-\frac{l_P^2\boldsymbol{\nabla}^2}{4}}\delta(\mathbf{r}-\boldsymbol{\xi})=
\delta(\mathbf{r}-\boldsymbol{\xi}) +  \left (e^{-\frac{l_P^2\boldsymbol
		{\nabla}^2}{2}}-e^{-\frac{l_P^2\boldsymbol{\nabla}^2}{4}}\right )
e^{\frac{l_P^2\boldsymbol{\nabla}^2}{4}}\,\delta(\mathbf{r}-\boldsymbol{\xi})~, $$
and recalling \eqref{eq4} we get
\begin{eqnarray} &&
e^{-\frac{l_P^2\boldsymbol{\nabla}^2}{4}}\delta(\mathbf{r}-\boldsymbol{\xi})=
\delta(\mathbf{r}-\boldsymbol{\xi})+\left (e^{-\frac{l_P^2\boldsymbol
		{\nabla}^2}{2}}-e^{-\frac{l_P^2\boldsymbol{\nabla}^2}{4}}\right )
f(\mathbf{r}-\boldsymbol{\xi})   = \nonumber \\ &&  \delta(\mathbf{r}-
\boldsymbol{\xi})+
\sum\limits_{n=1}^\infty\frac{(-l_P^2/4)^n}{n!}\,(2^n-1)\,\Delta^n
f(\mathbf{r}-\boldsymbol{\xi})~. 
\label{eq7}
\end{eqnarray}
Therefore, the inversion of  \eqref{Verschmierte Wellenfunktion} takes the form
\begin{eqnarray}
\Psi(\mathbf{r})=\widetilde{\Psi}(\mathbf{r})+  \sum\limits_{n=1}^\infty
\frac{(-l_P^2/4)^n}{n!}\,(2^n-1)\int \mathrm{d}^3\xi\,
\widetilde{\Psi}(\mathbf{r}-\boldsymbol{\xi})\,\Delta^n f(\boldsymbol{\xi}) ~. 
\label{eq8}
\end{eqnarray}
Derivatives of the Gaussian function \eqref{Gauss-Verteilung} can be expressed 
through the multivariate Hermite polynomials introduced by Grad 
\cite{Grad:1949}. One can use the definition 
\begin{equation}
\widetilde{H}^{(n)}_{i_1 i_2\ldots i_n}(\mathbf{r};l_P)=(-l_P^2)^n\,f^{-1} 
(\mathbf{r})\boldsymbol{\nabla}_{i_1}\boldsymbol
{\nabla}_{i_2}\cdots \boldsymbol{\nabla}_{i_n}f(\mathbf{r})~,
\label{eq9}  
\end{equation}
which generalizes the Rodrigues formula for the univariate Hermite 
polynomials \cite{Holmquist:1996} and simultaneously make multivariate 
Hermite polynomials dimensionless. Then \eqref{eq8} takes the form
\begin{eqnarray}
\Psi(\mathbf{r})=\widetilde{\Psi}(\mathbf{r})+\sum\limits_{n=1}^\infty
\frac{(-1)^n}{4^n\, n!}\,(2^n-1)\times   \int \mathrm{d}^3\xi\,
f(\boldsymbol{\xi})H_{2n}(\boldsymbol{\xi}^2/l_P^2)
\widetilde{\Psi}(\mathbf{r}-\boldsymbol{\xi}) ~,  
\label{eq10}
\end{eqnarray} 
where
\begin{equation}
H_{2n}(\boldsymbol{\xi}^2/l_P^2)=\delta_{i_1i_2}\delta_{i_3i_4}\cdots
\delta_{i_{2n-1}i_{2n}}\widetilde{H}^{(n)}_{i_1 i_2\ldots i_{2n-1}i_{2n}}
(\boldsymbol{\xi};l_P) ~,
\label{eq11}
\end{equation}
is completely contracted version of the multivariate Hermite polynomials
(a so called scalar irreducible Hermite polynomials \cite{Balescu:1988}).

Therefore, instead of \eqref{eq5}, the modified Schr\"{o}dinger equation can be 
written as
\begin{eqnarray} &&
i\hbar \partial_t \widetilde{\Psi}(\mathbf{r}) \,=\, -\frac{\hbar^2}{2m} 
\Delta \widetilde{\Psi}(\mathbf{r}) +
\int  \mathrm{d}^3\xi \, f(\mathbf{r}-\boldsymbol{\xi})V(\boldsymbol{\xi})
\widetilde{\Psi}(\boldsymbol{\xi})+
\sum\limits_{n=1}^\infty\frac{(-1)^n}{4^n\, n!}\,(2^n-1)\times \nonumber \\ &&
\iint \mathrm{d}^3\xi_1\,\mathrm{d}^3\xi_2\,f(\boldsymbol{\xi_1})\,
f(\boldsymbol{\xi_2})\,V(\mathbf{r}-\boldsymbol{\xi_1})
H_{2n}(\boldsymbol{\xi_2}^2/l_P^2)\widetilde{\Psi}(\mathbf{r}-
\boldsymbol{\xi_1}-\boldsymbol{\xi_2}).
\label{eq12}
\end{eqnarray}
Note that the scalar irreducible Hermite polynomials here can be expressed 
through Laguerre polynomials \cite{Balescu:1988,Lingam:2017}. Next point is 
to restore the Hermiticity of the Hamiltonian. 
However, we will not delve into these issues, despite the fact that the 
approach presented in this section is somewhat more general, since it is less 
important for our purposes.

\subsection{Perturbative corrections}
\label{shestsorebebi}
Let us list a few facts that immediately follow from the above discussion. 
First of all let us see how does a free particle wave-packet get modified
$$ \int\mathrm{d}^3k \, \mathrm{e}^{-i(\omega(\mathbf{k}) t - \mathbf{k}\cdot\mathbf{r})} 
g(\mathbf{k}) \, \to \, 
\int\mathrm{d}^3k \, \mathrm{e}^{-i(\omega(\mathbf{k}) t - \mathbf{k}\cdot\mathbf{r})} 
g(\mathbf{k})  
\int\mathrm{d}^3\xi \, f(\xi) \mathrm{e}^{-i\mathbf{k}\cdot \boldsymbol{\xi}} ~, $$
where $\omega(\mathbf{k}) = \hbar k^2/2m$. Denoting by 
$\tilde{f}(\mathbf{k})$ the Fourier transform 
of $f(\xi)$, one sees that the above modification amounts to replacing 
$g(\mathbf{k})$ by the product $g(\mathbf{k})
\tilde{f}(\mathbf{k}) \equiv \bar{g}(\mathbf{k})$. As the function 
$\tilde{f}(\mathbf{k})$ decays fast for $k \gtrsim k_P$, 
so does $\bar{g}(\mathbf{k})$. Thus, the result is that the wave-function 
can not be localized beneath the Planck length.          

When the particle moves in a potential field, for $V(\mathbf{r})$ and 
$\Psi(\mathbf{r})$ that vary negligibly over the Planck length, 
one can safely use the decompositions 
$$ V(\mathbf{r} - \boldsymbol{\xi}) \,=\, \sum_j \frac{(-\boldsymbol{\xi}\cdot
	\nabla)^j}{J!} V(\mathbf{r}) ~,  ~~ 
\Psi(\mathbf{r} - \boldsymbol{\xi}) \,=\, \sum_j \frac{(-\boldsymbol{\xi}
	\cdot\nabla)^j}{J!} \Psi(\mathbf{r}) ~, $$
and treat the equations (\ref{Grundgleichung}, 
\ref{einfache Version}) perturbatively. For Eq.\eqref{Grundgleichung} one 
obtains 
$$ i\hbar \partial_t \Psi(\mathbf{r}) \,=\, -\frac{\hbar^2}{2m} \Delta
\Psi(\mathbf{r}) \,+\, V(\mathbf{r})\Psi(\mathbf{r})  \,+\, $$ $$ 
\Big( \Psi\Delta V/4 \,+\,   \nabla \Psi \cdot \nabla V \,+\, V\Delta\Psi 
\Big) \frac{1}{6} \int\mathrm{d}^3\xi \, f(\xi)\xi^2  \,+\, 
\text{higher order terms}   ~. $$ 
It is plain to see that $\int\mathrm{d}^3\xi \, f(\xi)\xi^2$ 
is of the order of $l_P^2$. Correspondingly, the energy perturbations read
\begin{eqnarray*}
	\delta \mathcal{E}_j \,\propto\, l^2_P\int\mathrm{d}^3x \, \Psi^*_j 
	\Big( \Psi_j\Delta V/4 \,+\, \nabla \Psi_j \cdot \nabla 
	V \,+\, V\Delta\Psi_j \Big)  \\ = l^2_P\int\mathrm{d}^3x \, 
	|\Psi_j|^2 \Delta V/4 \,-\, l^2_P\int\mathrm{d}^3x \, 
	V|\nabla \Psi_j|^2 ~.
\end{eqnarray*} As usual, $\Psi_j$ functions are assumed to be normalized and 
orthogonal to one another. In the case of 
Eq.\eqref{einfache Version}, the energy corrections take the form  
\begin{eqnarray*}
	\delta \mathcal{E}_j \,\propto\, l^2_P\int\mathrm{d}^3x \, |\Psi_j|^2 
	\Delta V ~.  
\end{eqnarray*} One sees that, in general, the corrections to the 
energy eigenvalues are real. 

\subsection{Modified Schr\"{o}dinger equation and equivalence principle}
\label{Equivalence_Principle}
It is interesting to consider a perturbative analysis of the modified 
Schr\"{o}dinger equation in the case of a one-dimensional linear potential 
$V(x)=mgx$, assuming that the smearing function is Gaussian and is given by
the formula $f(\xi)=(\pi l_P^2)^{-1/2}e^{-\xi^2/l_P^2}$. It is clear that for 
Eq.\eqref{einfache Version} there are no corrections at all, 
so we will concentrate on the case of Eq.\eqref{Grundgleichung}. For the 
one-dimensional linear potential, it has the form
\begin{equation}
i\hbar \,\frac{\partial \Psi(x,t)}{\partial t} \,=\, -\frac{\hbar^2}{2m} \frac{ 
\partial ^2 \Psi(x,t)}{\partial x^2} \,+\,  mg\int\limits_{-\infty}^\infty d\xi \, 
f(\xi)\left(x-\frac{\xi}{2}\right)\Psi(x-\xi,t) ~. 
\label{EP1}
\end{equation}
Due to the presence of the smearing function $f(\xi)$, essentially only the
region $|x-\xi|\le l_P$ contributes to the integral. Hence we expand
$$\Psi(x-\xi,t)\approx \Psi(x,t)-\frac{\partial \Psi(x,t)}{\partial x}\,\xi+
\frac{1}{2}\frac{\partial^2 \Psi(x,t)}{\partial x^2}\,\xi^2,$$
and taking into account that $\int\limits_{-\infty}^\infty d\xi \, 
f(\xi)=1$ and $\int\limits_{-\infty}^\infty d\xi \, f(\xi)\,\xi^2=l_P^2/2$,
we end up with the equation
\begin{equation}
i\hbar \,\frac{\partial \Psi(x,t)}{\partial t} \,=\, -\frac{\hbar^2}{2m} \frac{ 
\partial ^2 \Psi(x,t)}{\partial x^2} \,+\, mgx\,\Psi(x,t)\,+\,
\frac{mgl_P^2}{4}\left [x\,\frac{\partial ^2 \Psi(x,t)}{\partial x^2}+
\frac{\partial \Psi(x,t)}{\partial x}\right ] ~. 
\label{EP2}
\end{equation}
It is convenient to rewrite Eq.{\eqref{EP2}} in the form
\begin{equation}
i\hbar \,\frac{\partial \Psi(x,t)}{\partial t} \,=\, \left [\frac{\hat{p}^2}
{2m}\,+\, mgx\,-\,\frac{mgl_P^2}{4\hbar^2}\left (x\hat{p}^2-i\hbar\hat p
\right )\right ]\Psi(x,t) ~, 
\label{EP3}
\end{equation}
where $\hat p=-i\hbar\frac{\partial}{\partial x}$ is the momentum operator in
the coordinate representation. But
$$i\hbar\hat p=\frac{1}{2}\left (\hat p\,[x,\hat p]+[x,\hat p]\,\hat p\right )=
\frac{1}{2}\left (x\hat{p}^2-\hat{p}^2x\right ),$$
and \eqref{EP3} is equivalent to
\begin{equation}
i\hbar \,\frac{\partial \Psi(x,t)}{\partial t} \,=\, \left [\frac{\hat{p}^2}
{2m}\,+\, mgx\,-\,\frac{mgl_P^2}{8\hbar^2}\left (x\hat{p}^2+\hat{p}^2 x
\right )\right ]\Psi(x,t) ~, 
\label{EP4}
\end{equation}
from which it is clear that the Hamiltonian is Hermitian.

The equation \eqref{EP4} has a formal solution
\begin{equation}
\Psi(x,t)=e^{-\frac{it}{\hbar}\left [\frac{\hat{p}^2}
{2m}\,+\, mgx\,-\,\frac{mgl_P^2}{8\hbar^2}\left (x\hat{p}^2+\hat{p}^2 x
\right )\right ]}\Psi(x,0).
\label{EP5}
\end{equation}
To turn this formal solution into a true solution, we must disentangle 
non-commutative operators in this formal solution. This can be done using the
left-oriented Zassenhaus formula \cite{Zassenhaus1,Zassenhaus2,Zassenhaus3}
\begin{equation}
e^{\lambda\left (\hat X+\hat Y\right )}=\cdots e^{\lambda^5\hat C_5(\hat X,\hat Y)}
e^{\lambda^4\hat C_4(\hat X,\hat Y)}e^{\lambda^3\hat C_3(\hat X,\hat Y)}
e^{\lambda^2\hat C_2(\hat X,\hat Y)}e^{\lambda\hat Y}e^{\lambda\hat X},
\label{EP6}
\end{equation}
where \cite{Zassenhaus2,Zassenhaus3}
\begin{equation} 
\begin{aligned}
& \hat C_2(\hat X,\hat Y)=\frac{1}{2}[\hat X,\,\hat Y],\;\;\;
\hat C_3(\hat X,\hat Y)=\frac{1}{3}[\hat Y,\,[\hat X,\,\hat Y]]+
\frac{1}{6}[\hat X,\,[\hat X,\,\hat Y]], \\
& \hat C_4(\hat X,\hat Y)=\frac{1}{8}\left ([\hat Y,\,[\hat Y,\,[
\hat X,\,\hat Y]]]+[\hat Y,\,[\hat X,\,[\hat X,\,\hat Y]]]\right )+
\frac{1}{24}[\hat X,\,[\hat X,\,[\hat X,\,\hat Y]]], \\
&\hat C_5(\hat X,\hat Y)=\frac{1}{30}\left ([\hat Y,\,[\hat Y,\,[\hat Y,\,
[\hat X,\,\hat Y]]]]+[\hat Y,\,[\hat X,\,[\hat X,\,[\hat X,\,\hat Y]]]]
\right )+\frac{1}{10}[[[\hat X,\,\hat Y],\,\hat Y],\,
[\hat X,\,\hat Y]]\\
&+\frac{1}{20}\left ([\hat Y,\,[\hat Y,\,[\hat X,\,
[\hat X,\,\hat Y]]]]+[[[\hat X,\,\hat Y],\,\hat X],\,[\hat X,\,\hat Y]]
\right )+\frac{1}{120}[\hat X,\,[\hat X,\,[\hat X,\,
[\hat X,\,\hat Y]]]]\, .
\end{aligned}
\label{EP7}
\end{equation}
In our case we can take
\begin{equation}
\lambda=-\frac{it}{2m\hbar},\;\;\;\hat X=\hat{p}^2+2m^2gx,\;\;\;
\hat Y=-\frac{m^2gl_P^2}{4\hbar^2}\left (x\hat{p}^2+\hat{p}^2x\right ),
\label{EP8}
\end{equation}
and calculate non-zero nested commutators (only terms $\sim l_P^2$ are 
retained)
\begin{equation} 
\begin{aligned}
& \hat C_2(\hat X,\hat Y)=\frac{1}{2}[\hat X,\,\hat Y]
=\frac{im^2gl_P^2}{2\hbar}\left [\hat{p}^3-m^2g\left(x\hat{p}^2+
\hat{p}^2x\right)\right],\\
&\hat C_3(\hat X,\hat Y)\approx\frac{1}{6}[\hat X,\,[\hat X,\,\hat Y]]
=-\frac{1}{3}m^4g^2l_P^2\left (5\hat{p}^2-2m^2gx\right ), \\
& \hat C_4(\hat X,\hat Y)\approx \frac{1}{24}[\hat X,\,[\hat X,\,[\hat X,\,
\hat Y]]]=-2i\hbar m^6g^3l_P^2\,\hat{p},\\
&\hat C_5(\hat X,\hat Y)\approx \frac{1}{120}[\hat X,\,[\hat X,\,[\hat X,\,
[\hat X,\,\hat Y]]]]=\frac{4}{5}\hbar^2m^8g^4l_P^2\,.
\end{aligned}
\label{EP9}
\end{equation}
Then
\begin{eqnarray} & 
e^{\lambda^5\hat C_5(\hat X,\hat Y)}e^{\lambda^4\hat C_4(\hat X,\hat Y)}
e^{\lambda^3\hat C_3(\hat X,\hat Y)}e^{\lambda^2\hat C_2(\hat X,\hat Y)}e^{\lambda\hat Y}\approx 
& \nonumber \\ & 1+\lambda\hat Y+
\lambda^2\hat C_2(\hat X,\hat Y)+\lambda^3\hat C_3(\hat X,\hat Y)+
\lambda^4\hat C_4(\hat X,\hat Y)+\lambda^5\hat C_5(\hat X,\hat Y)=
1+\hat F(x,\hat p), &
\label{EP10}
\end{eqnarray}
where
\begin{equation}
\begin{aligned}
& \hat F(x,\hat p)=\left . \frac{igtl_P^2}{8\hbar^3}\right [m\left (x\hat{p}^2
+\hat{p}^2x\right )-t\left [\hat{p}^3-m^2g\left (x\hat{p}+\hat{p}x
\right )\right ]- \\
& \left . \frac{t^2}{3}\,mg\left (5\hat{p}^2-2m^2gx\right )
-t^3m^2g^2\hat{p}-\frac{t^4}{5}\,m^3g^3\right ].
\end{aligned}
\label{EP11}
\end{equation}
On the other hand, $\hat X=\hat A+\hat B$ with $\hat A=
{\hat p}^2$, $\hat B=2m^2gx$, and with the following non-zero nested
commutators:
\begin{equation}
[\hat A,\,\hat B]=-4i\hbar m^2g\hat p,\;\;[\hat A,\,[\hat A,\,\hat B]]=0,\;\;
[\hat B,\,[\hat A,\,\hat B]]=8\hbar^2 m^4g^2.
\label{EP12}
\end{equation}
Therefore, again using the Zassenhaus formula, we get
\begin{equation}
e^{\lambda\hat X}\Psi(x,0)=e^{\frac{i}{\hbar}\frac{mg^2t^3}{3}}\,e^{\frac{i}{\hbar}\frac{gt^2}{2}\,
\hat p}e^{-\frac{i}{\hbar}\,mg\hat x t}e^{-\frac{i}{\hbar}\,\frac{{\hat p}^2}
{2m}\,t}\,\Psi (x,0).
\label{EP13}
\end{equation}
But $e^{-\frac{i}{\hbar}\,\frac{{\hat p}^2}{2m}\,t}\,\Psi (x,0)=\Psi_{\mathrm{free}}(x,t)$
gives  the wave function of a free particle - a solution of the 
Schr\"{o}dinger equation with zero potential, and $e^{\frac{i}{\hbar}a\hat p}$ 
for any real number $a$ is a spatial translation operator:
$$e^{\frac{i}{\hbar}\frac{gt^2}{2}\,\hat p}e^{-\frac{i}{\hbar}\,mg\hat x t}\,
\Psi_{\mathrm{free}}(x,t)=e^{-\frac{i}{\hbar}\,mg\left(x+\frac{gt^2}{2}\right) t}\,
\Psi_{\mathrm{free}}\left (x+\frac{gt^2}{2},
t\right ).$$
Therefore,
\begin{equation}
e^{\lambda\hat X}\Psi(x,0)=e^{-\frac {imgt}{\hbar }\left( x+
\frac {gt^{2}}{6}\right)}\,\Psi_{\mathrm{free}} 
\left( x+\frac {gt^{2}}{2},t\right),
\label{EP14}
\end{equation}
and finally \eqref{EP5} takes the form
\begin{equation}
\Psi(x,t)=\left [1+\hat F(x,\hat p)\right ] e^{-\frac {imgt}{\hbar }\left( x+
\frac {gt^{2}}{6}\right)}\,\Psi_{\mathrm{free}} 
\left( x+\frac {gt^{2}}{2},t\right).
\label{EP15}
\end{equation} 
As explained in \cite{QEP1,QEP2}, without $\hat F(x,\hat p)$ term, this
relation constitutes a quantum-mechanical embodiment of  Einstein's principle 
of equivalence. The presence of the $\hat F(x,\hat p)$ term indicates that
the equivalence principle is violated. However, the violation is minuscule
and beyond the experimental reach, for example, in neutron quantum bouncing 
experiments in the Earth's gravitational field  \cite{QEP3}. It is convenient 
to introduce the time $\tau=c/g\approx 1\mathrm{year}$ and express 
$\hat F(x,\hat p)$ in dimensionless units:
\begin{equation}
\begin{aligned}
& \hat F(x,\hat p)=\left . i\,\frac{l_P^2m^3c^5}{8\hbar^3 g}\right [
\frac{g\left (x\hat{p}^2+\hat{p}^2x\right )}{m^2c^4}\,\frac{t}{\tau}-
\frac{\hat{p}^3-m^2g\left (x\hat{p}+\hat{p}x
\right )}{m^3c^3}\,\left (\frac{t}{\tau}\right )^2- \\
& \left . \frac{1}{3}\,\frac{5\hat{p}^2-2m^2gx}{m^2c^2}\,
\left (\frac{t}{\tau}\right )^3-\frac{\hat{p}}{mc}\,
\left (\frac{t}{\tau}\right )^4
-\frac{1}{5}\,\left (\frac{t}{\tau}\right )^5\right ].
\end{aligned}
\label{EP16}
\end{equation} 
For neutron, $\frac{l_P^2m^3c^5}{8\hbar^3 g}\approx 3\cdot 10^{-8}$.

The non-local corrections of the Schr\"{o}dinger equation are intrinsically 
linked to the Planck length similarly to GUP. It is not surprising, therefore,
that similar conclusions about a slight violation of the equivalence principle 
were made in \cite{QEP4} considering GUP corrections to the geodesic equation,
and in \cite{7D} when considering Lorentz invariant length scale.

\subsection{Semi-classical limit}
\label{klasika}

In the case of Eq.\eqref{einfache Version}, the discussion of the 
semi-classical limit is straightforward. From now on 
let us assume Gaussian fluctuations for the background space 
\eqref{Gauss-Verteilung}. Then one can evaluate the 
integral defining a non-local term in Eq.\eqref{Grundgleichung} as follows 
(see ref. \cite{1962NucPh..32..353P}) 
\begin{eqnarray*}
	i\hbar \partial_t \Psi(\mathbf{r}) \,=\, -\frac{\hbar^2}{2m} \Delta 
	\Psi(\mathbf{r}) \,+\,  \exp\left(\frac{l_P^2}{4} 
	[\nabla_1/2 + \nabla_2]^2\right) V(\mathbf{r})\Psi(\mathbf{r}) ~,   
\end{eqnarray*} where $\nabla_1$ acts on $V$ and $\nabla_2$ on $\Psi$, 
respectively. Once again, one sees that if 
$V$ and $\Psi$ vary slowly over the distance $l_P$, the corrections are 
strongly suppressed. The WKB approximation 
to the integro-differential equation \eqref{Grundgleichung} has been discussed 
in \cite{Horiuchi:1980hz}. For our purposes 
it is expedient to write the Eq. \eqref{Grundgleichung} in the form 
\cite{Horiuchi:1980hz}      
\begin{eqnarray*}
i\hbar \partial_t \Psi \,=\, \left\{ \frac{\widehat{\mathbf{p}}^2}{2m} \,+\, 
\int\mathrm{d}^3\xi \, f(\xi) \, 
\mathrm{e}^{-i\boldsymbol{\xi}\cdot\widehat{\mathbf{p}}/2\hbar} V\left(\mathbf{r}\right)
	\mathrm{e}^{-i\boldsymbol{\xi}\cdot\widehat{\mathbf{p}}/2\hbar} \right\} \Psi ~.  
\end{eqnarray*} Derivation of the Heisenberg equations can be safely 
accomplished by allowing operators to act on 
a wave-function, which is removed at the end of calculation. Doing it in the 
coordinate representation, one obtains 
\begin{eqnarray}\label{modzganterti} &&
\dot{\widehat{\mathbf{r}}} \, \Psi(\mathbf{r}) \,=\, i\left[\widehat{H}, 
\widehat{\mathbf{r}}\right]\, \Psi(\mathbf{r}) \,=\, i 
\left[ \frac{\widehat{\mathbf{p}}^2}{2m}, \widehat{\mathbf{r}} \right]
\Psi(\mathbf{r}) \,+\, i \int\mathrm{d}^3x' 
\,(\mathbf{r}' - \mathbf{r}) f(|\mathbf{r} - \mathbf{r}'|)
\times \nonumber \\ &&
V\left(\frac{\mathbf{r} +\mathbf{r}'}{2}\right)\Psi(\mathbf{r}') 
\,=\,  \frac{\widehat{\mathbf{p}}}{m} \, \Psi(\mathbf{r}) \,-\,
i \int\mathrm{d}^3\xi \, \boldsymbol{\xi} f(\xi) 
V\left(\mathbf{r} - \boldsymbol{\xi}/2\right) \Psi(\mathbf{r} - 
\boldsymbol{\xi} ) \,= \nonumber \\ &&
\frac{\widehat{\mathbf{p}}}{m} \, \Psi(\mathbf{r}) \,-
i \int\mathrm{d}^3\xi \, \boldsymbol{\xi} f(\xi) \, 
\mathrm{e}^{-i\boldsymbol{\xi}\cdot\widehat{\mathbf{p}}/2\hbar} V\left(\mathbf{r} \right) 
\mathrm{e}^{-i\boldsymbol{\xi}
	\cdot\widehat{\mathbf{p}}/2\hbar}\,\Psi(\mathbf{r}) ~,  
\end{eqnarray} 
and 
\begin{eqnarray}\label{modzgantori} &&
\dot{\widehat{\mathbf{p}}} \, 
\Psi(\mathbf{r}) \,=\, \int\mathrm{d}^3x' \, f(|\mathbf{r} - \mathbf{r}'|)
V\left(\frac{\mathbf{r} +\mathbf{r}'}{2}\right)
\nabla_{\mathbf{r}'}\Psi(\mathbf{r}') \,-\,
\nabla_{\mathbf{r}} \int\mathrm{d}^3x' \, f(|\mathbf{r} - \mathbf{r}'|)
\times \nonumber \\ &&
V\left(\frac{\mathbf{r} +\mathbf{r}'}{2}\right)\Psi(\mathbf{r}')  
\,=\,  -2  \int\mathrm{d}^3x' \, f(|\mathbf{r} - 
\mathbf{r}'|)\nabla_{\mathbf{r}}V\left(\frac{\mathbf{r} +\mathbf{r}'}{2}\right)
\Psi(\mathbf{r}') \,=\nonumber \\ && 
- \int\mathrm{d}^3\xi \, f(\xi) \, 
\mathrm{e}^{-i\boldsymbol{\xi}\cdot\widehat{\mathbf{p}}/2\hbar} 
\nabla_{\mathbf{r}} V\left(\mathbf{r} \right) 
\mathrm{e}^{-i\boldsymbol{\xi}\cdot\widehat{\mathbf{p}}/2\hbar} 
\Psi(\mathbf{r}) ~.
\end{eqnarray} Thus, the Heisenberg equations (\ref{modzganterti}, 
\ref{modzgantori}) read 
\begin{eqnarray*} &&
\dot{\widehat{\mathbf{p}}} \,=\, - \int\mathrm{d}^3\xi \, f(\xi) \, 
\mathrm{e}^{-i\boldsymbol{\xi}\cdot\widehat{\mathbf{p}}/2\hbar} \nabla V\left(\mathbf{r} 
\right) \mathrm{e}^{-i\boldsymbol{\xi}\cdot\widehat{\mathbf{p}}/2\hbar} ~, \\&& 
\dot{\widehat{\mathbf{r}}} \,=\, \frac{\widehat{\mathbf{p}}}{m} \,-\, i 
\int\mathrm{d}^3\xi \, \boldsymbol{\xi} f(\xi) \, 
\mathrm{e}^{-i\boldsymbol{\xi}\cdot\widehat{\mathbf{p}}/2\hbar} 
V\left(\mathbf{r} \right) \mathrm{e}^{-i\boldsymbol{\xi}\cdot\widehat{\mathbf{p}}/2\hbar} ~.
\end{eqnarray*} As to the equations of classical motion, they can be written 
immediately by using the modified Hamiltonian    
\begin{eqnarray} &&
\mathcal{H} \,=\, \frac{\mathbf{p}^2}{2m} \,+\, \int\mathrm{d}^3\xi \, 
f(\xi) \mathrm{e}^{-i\boldsymbol{\xi}
	\cdot\mathbf{p}/2\hbar} V\left(\mathbf{r}\right)
\mathrm{e}^{-i\boldsymbol{\xi}\cdot\mathbf{p}/2\hbar}  \,= \nonumber \\ &&  
\frac{\mathbf{p}^2}{2m} \,+\, V\left(\mathbf{r}\right) 
\exp\left(-\frac{l_P^2\mathbf{p}^2}{4\hbar^2}\right)  ~,  \end{eqnarray}
\noindent which gives (these equations have already been discussed in 
\cite{1976ZPhyA.276...79T,1980PThPh..63..725H})
\begin{eqnarray*}
\dot{\mathbf{r}} \,=\, \frac{\mathbf{p}}{m} \,-\, 
\frac{l_P^2 V\left(\mathbf{r}\right)\mathbf{p}}{2\hbar^2}  \exp
\left(-\frac{l_P^2\mathbf{p}^2}{4\hbar^2}\right) ~, ~~~~ \dot{\mathbf{p}} 
\,=\, - \, \nabla V\left(\mathbf{r}\right) 
\exp\left(-\frac{l_P^2\mathbf{p}^2}{4\hbar^2}\right) ~.  
\end{eqnarray*} The deviation from the standard dynamics disappears as long as 
the condition $ \mathbf{p}^2 \ll \hbar^2/l_P^2$ is fulfilled. That means that 
one should require 
\begin{eqnarray*}
\mathcal{E} \,-\, V\left(\mathbf{r}\right) \, \ll \, \frac{\hbar^2}{ml_P^2} ~, 
\nonumber \end{eqnarray*} 
where $\mathcal{E}$ stands for energy. When we are 
dealing with the classical motion of macroscopic objects, 
this requirement is often broken and we face the above mentioned soccer-ball 
problem. For example the earth has average orbital 
speed $\approx 30$ km/s and the mass $\approx 6\times 10^{24}$ kg while 
$\hbar / l_P \approx 6.5 $ kg$\cdot$m /s. 
In this particular case the condition $\mathbf{p}^2 \gg \hbar^2/l_P^2$ is 
satisfied extremely well. In view of the modified 
dynamics, it implies that with a great accuracy $\dot{\mathbf{r}} = 
\mathbf{p}/m$ and 
\begin{eqnarray*}
	\dot{\mathbf{p}} = - \, \nabla V\left(\mathbf{r}\right)
	\exp\left(-\frac{l_P^2\mathbf{p}^2}{4\hbar^2}\right) ~.  
\end{eqnarray*} Taking into account that the exponential factor is in this case 
of the order of $\exp\left(-10^{55}\right)$, 
the motion of earth around the sun should be drastically altered. One possible 
solution to this dramatic puzzle was 
considered in \cite{Quesne,Tkachuk,Gnatenko} and suggests that the effective 
Planck length for composite objects is many orders of magnitude less than $l_P$.

Corrections to the classical dynamics implied by the 
Eq.\eqref{einfache Version} is of course harmless. Namely, in this case the 
corrections arise due to modification of the potential 
$$ V(\mathbf{r}) \,\rightarrow \, \int\mathrm{d}^3\xi \, 
\frac{\mathrm{e}^{-\xi^2/l_P^2} }{\pi^{3/2}l_P^3} \, V(\mathbf{r} - 
\boldsymbol{\xi}) \,=\, \exp\left(\frac{l_P^2\Delta}{4}\right) V(\mathbf{r})  
\,=$$ $$  V(\mathbf{r}) \,+\, \frac{l_P^2\Delta}{4}  
V(\mathbf{r}) \,+\, \cdots ~. $$
One could again consider an orbit of the earth and calculate in 
particular a perihelion shift but for the potential 
$\propto r^{-1}$ there are no corrections as $\Delta r^{-1} = 0$ for 
$\mathbf{r}\neq \mathbf{0}$. Moreover, one can claim that,
in general, the modified theory given by Eq.\eqref{einfache Version} should 
not affect the classical regime. To see it, let us note that 
the Hamiltonian in this case can be written as
\begin{eqnarray*}
\mathcal{H} \,=\, \frac{\mathbf{p}^2}{2m} \,+\, \int\mathrm{d}^3\xi \, f(\xi) 
\mathrm{e}^{-i\boldsymbol{\xi}\cdot\widehat{\mathbf{p}}/\hbar} V\left(\mathbf{r}\right)
\mathrm{e}^{i\boldsymbol{\xi}\cdot\widehat{\mathbf{p}}/\hbar} ~,  
\end{eqnarray*} and, therefore, one arrives at the standard Hamiltonian in 
classical regime, since in this regime the momentum and position operators do 
commute.

\section{Concluding remarks}

The general reasoning so far given can readily be compared with the momentum 
cut-off approach for implementing the concept of minimum length into QM 
\cite{Sailer:2013yk,Sailer:2014hea}. This approach implies to restrict the 
Hilbert space of state vectors to the cut-off functions 
\begin{eqnarray*}
	\Psi(\mathbf{r}) \,=\,  \int\limits_{k<k_P} \mathrm{d}^3k \, 
	\mathrm{e}^{-i\mathbf{k}\cdot\mathbf{r}} \chi(\mathbf{k}) ~ , 
\end{eqnarray*} where $k_P$ stands for the Planck momentum: 
$k_P = \sqrt{ c^3 /\hbar G_N}$. The averaging in 
Eq.\eqref{Verschmierte Wellenfunktion} does basically the same job. 
The approach based on 
Eq.\eqref{Verschmierte Wellenfunktion} for deriving the modified 
Schr\"{o}dinger equation may be somewhat advantageous
in treating the product $V(\mathbf{r})\Psi(\mathbf{r})$. An advantage of the 
approach based on Eq.\eqref{Verschmierte 
	Wellenfunktion} is that it guides logically in treating the product 
$V(\mathbf{r})\Psi(\mathbf{r})$. Also it makes easy to work 
out the corrections to QM and address the question of a classical limit. 

Apart from the trivial generalization given by Eq.\eqref{einfache Version}, we 
see that the non-local theory, when naively 
applied to composite objects, leads to unacceptably large effects in the 
classical limit. Thus, we face the same impasse as in the 
case of deformed Weyl-Heisenberg algebra 
\cite{Silagadze:2009vu,Maziashvili:2012zr}. As we have seen, the classical 
limit of band-limited QM also suffers from this soccer-ball problem. To date, 
this problem has not been satisfactorily resolved, and there 
is no universally accepted solution to this problem \cite{Sabina}.

Interestingly, we are now in a position to write down the minimum-length 
modified QM that in principle has good semi-classical 
behavior. Indeed, one may attempt to restore the standard classical picture 
for non-local theory by incorporating both above considered  modifications in 
a single equation 
$$i\hbar \partial_t \Psi(\mathbf{r}) \,=\, -\frac{\hbar^2}{2m} \Delta 
\Psi(\mathbf{r}) \,+ $$ $$  \int\mathrm{d}^3x' \, f(|\mathbf{r} - 
\mathbf{r}'|)V\left(\frac{\mathbf{r} +\mathbf{r}'}{2}\right) 
\Big\{\mathrm{w}_1\Psi(\mathbf{r}') + \mathrm{w}_2
\Psi(\mathbf{r})\Big\} ~, $$ 
where the weights, $0\leq \mathrm{w}_j \leq 1$, obey the 
relation $\mathrm{w}_1+\mathrm{w}_2=1$. That is, 
$\Psi(\mathbf{r}')$ and $\Psi(\mathbf{r})$ do not necessarily enter this 
equation with equal weights. In view of our conceptual 
framework given in section \ref{shesavali}, it is natural to assume that 
the effect of background space fluctuations should depend 
on the breadth of a wave-function as it determines the length scale probed 
by the particle. Similar considerations for the harmonic 
oscillator can indeed be used for estimating the rate of effect 
\cite{Maziashvili:2016kad}. Following this reasoning, by introducing 
\begin{eqnarray*}
	l^2 \,=\, \langle \Psi| \Big( \hat{\mathbf{r}} \,-\,  \langle \Psi|  
	\hat{\mathbf{r}}|\Psi\rangle\Big)^2|\Psi\rangle ~,  
\end{eqnarray*} as a standard measure of the spread of the wave-function, one 
could set the weights as $\mathrm{w}_1 = l_P/l$ 
to some power and $\mathrm{w}_2= 1- \mathrm{w}_1$. Equally well, for setting 
the weights one could use some other effective  
scale instead of $l_P$. If the breadth of the initial state is macroscopic, 
then $\mathrm{w}_1 \ll 1$, and one can safely omit the 
corresponding term, which will lead to the good classical behavior.  However, 
we are afraid, it is difficult to see how this rather ad-hoc solution of the 
soccer-ball problem could follow from the more fundamental theory.

\begin{acknowledgments}
This work has benefited from conversations with Zurab Kepuladze. We are also 
indebted to Hans-Thomas Elze for useful e-mail correspondences. The work of 
Z.K.S. is supported by the Ministry of Education and  Science of the Russian 
Federation.
\end{acknowledgments}



\end{document}